\documentclass[aps,prl,twocolumn,superscriptaddress,showpacs]{revtex4}

\usepackage{graphicx}
\usepackage{amsmath}
\usepackage{amssymb}
\usepackage{mathrsfs}
\usepackage{subfigure}
\usepackage{color}
\usepackage{bm}
\usepackage{hyperref}

\DeclareMathOperator{\trace}{tr}
\DeclareMathOperator{\sinc}{sinc}

\begin{document}

\author{Michael Foss-Feig}
\affiliation{JILA, NIST, and Department of Physics, University of Colorado, Boulder, CO 80309-0440, USA}

\author{Kaden R. A. Hazzard}
\affiliation{JILA, NIST, and Department of Physics, University of Colorado, Boulder, CO 80309-0440, USA}

\author{John J. Bollinger}
\affiliation{National Institute of Standards and Technology, Boulder, Colorado 80305, USA}

\author{Ana Maria Rey}
\affiliation{JILA, NIST, and Department of Physics, University of Colorado, Boulder, CO 80309-0440, USA}

\title{Non-equilibrium dynamics of Ising models with decoherence: an
  exact solution}

\begin{abstract}
The interplay between interactions and decoherence in
many-body systems is of fundamental importance in quantum physics:
Decoherence can degrade correlations, but can also give rise to a
variety of rich dynamical and steady-state behaviors.  We obtain an exact analytic solution for
the non-equilibrium dynamics of Ising models with arbitrary
interactions and subject to the most general form of local Markovian decoherence.  Our solution shows that
decoherence affects the relaxation of observables more than predicted by single-particle
considerations.  It also reveals a dynamical phase transition,
specifically a Hopf bifurcation, which is absent at the single-particle level.  These calculations
are applicable to ongoing quantum information and
emulation efforts using a variety of atomic, molecular, optical, and solid-state systems.
\end{abstract}

\maketitle


Understanding strongly correlated quantum systems in the presence of
decoherence is a fundamental challenge in modern physics.
While decoherence generally tends to degrade correlations, it is now
widely appreciated that it can also give rise to many-body physics not possible
with strictly coherent dynamics \cite{diehl1,torre,lee,PhysRevA.85.043620,PhysRevLett.109.020403,bs}, and can be used explicitly for the creation
of entanglement\cite{PhysRevLett.88.197901,kraus,diehl2,fossfeig}.  Regardless of whether
one's intention is to minimize or to harness decoherence,
determining its effect on interacting many-body systems is central to quantum simulation\cite{blochrmp},
quantum information \cite{horodecki}, and quantum metrology
\cite{giovannetti}.  So far, this understanding has been hindered by
the computational complexity of numerical techniques for open systems and the scarcity of exact
analytic solutions.  Exact solutions for dynamics of interacting
quantum systems in dimensions greater than one are rare even in the
absence of decoherence, and to our knowledge no such solutions have
been obtained in the presence of local decoherence.

The central result of this manuscript is an exact solution, Eqs. (\ref{tsl}-\ref{correlations2}), for the
time-dependence of all two-spin correlation functions in a system of
spins interacting via arbitrary Ising couplings, and subject to the most general form of
local Markovian decoherence allowed by nature \cite{carmichael}.  Our solution is applicable to a broad range of
important quantum systems, including trapped ions \cite{britton,monroe1,monroe2}, polar molecules \cite{kkni,gorshkov1},
Rydberg atoms \cite{weimer,low}, neutral atoms in
optical cavities \cite{sarang1,sachdev}, optical lattice
clocks \cite{swallows}, superconducting qubits\cite{mooij}, quantum dots\cite{hanson}, and nitrogen
vacancy centers\cite{prawer}.  Here we apply our solution to
trapped ion experiments because: (1) the relative importance of decoherence
and coherently driven quantum correlations is controllable, (2) the tunable long-ranged
interactions are generically frustrated, making large-scale numerical
simulations impractical, and (3) these experiments are the most
developed, with non-equilibrium dynamics already being explored \cite{britton}.


\emph{Outline of the calculation}.---We consider
far-from-equilibrium dynamics of a long-ranged Ising Hamiltonian
\begin{equation}
\label{hamiltonian}
\mathcal{H}=\frac{1}{\mathcal{N}}\sum_{i<j}J_{ij}\hat{\sigma}_i^z\hat{\sigma}_j^z.
\end{equation}
\begin{figure}[t!]
\includegraphics[width=8.5cm]{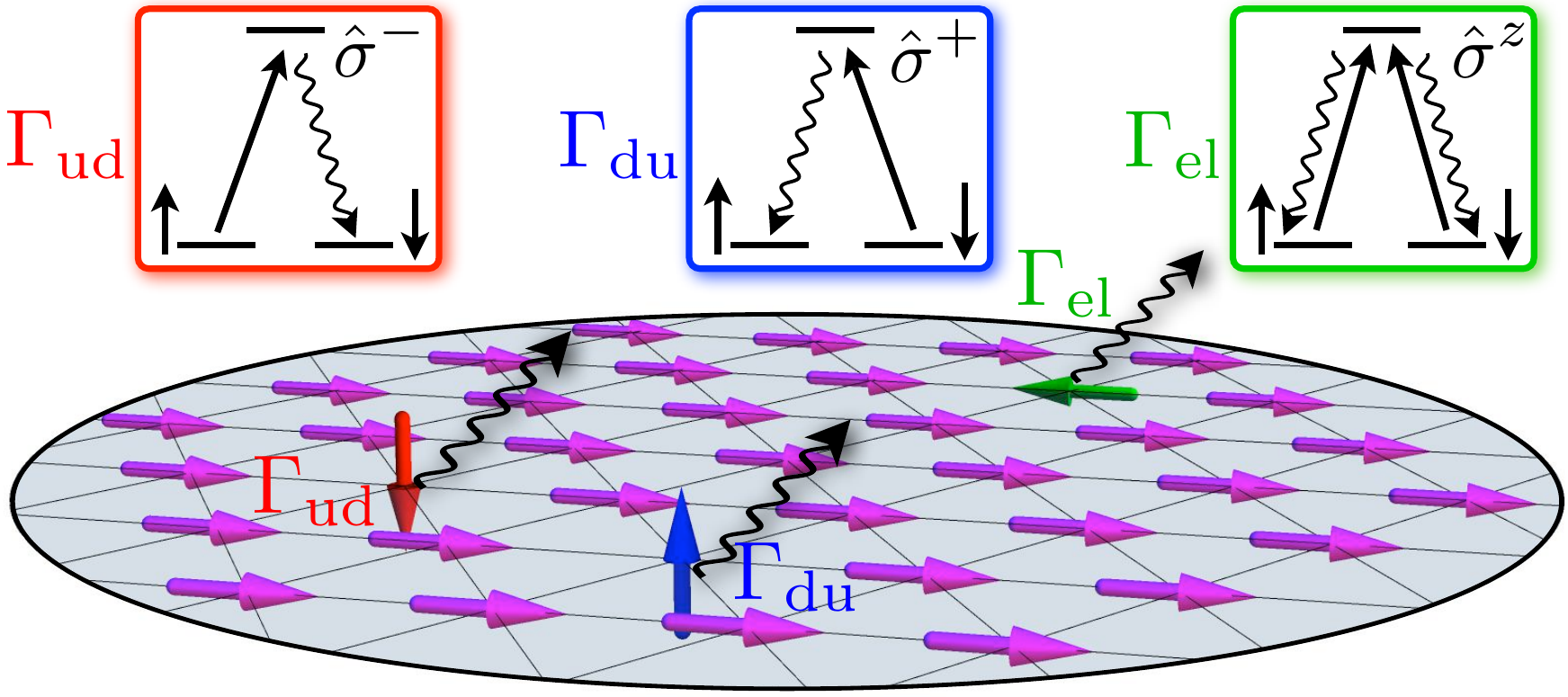}
\vspace{0.0cm}
\caption{ (Color online) Schematic illustration of the various
  decoherence processes.  The figure shows a lattice of spins
  initialized to point in a direction orthogonal to the $z$-axis.
  Longitudinal ($T_1$) spin relaxation with rates
  $\Gamma_{\mathrm{ud}},~\Gamma_{\mathrm{du}}$ (red and blue spin,
  respectively), and dephasing ($T_2$) with rate $\Gamma_{el}$ (green
  spin) are shown for three different spins.  In atomic systems,
  one way this decoherence can arise is due to spontaneous emission from an
  excited level (top panels) \cite{uys}.}
\label{Fig1}
\end{figure}
Here $\hat{\sigma}^a$ ($a=x,y,z$) are Pauli matrices, $\mathcal{N}$ is the total
number of spins, and Latin letters from the middle of the alphabet are
site indices.  Our results are valid for arbitrary $J_{ij}$, but given the relevance to numerous experiments we will
sometimes consider power law couplings $J_{ij}=J|\bm{r}_i-\bm{r}_j|^{-\zeta}$, where $\bm{r}_i$ is the
position of the $i^{\mathrm{th}}$ spin in lattice units ($\zeta=3$ for polar
molecules, $\zeta=6$ for Rydberg atoms, and $0<\zeta<3$ for trapped ions).  In the
presence of local decoherence, the most general Markovian dynamics of
the system reduced density matrix obeys a master equation
\footnote{In its current form it can
  describe, for instance, coupling of the spins to a thermal bath. The master equation could be slightly generalized by
  allowing the decoherence rates to vary from site to site.  This can
  be taken into account trivially by adding site indices to these
  rates in the final expressions,
  Eqs. (\ref{tsl}-\ref{correlations2}).}
\begin{equation}
\hbar\dot{\rho}=-i\left(\mathcal{H}_{\mathrm{eff}}\rho-\rho\mathcal{H}_{\mathrm{eff}}^{\dagger}\right)+\mathcal{D}(\rho).
\label{meq}
\end{equation}
The effective Hamiltonian $\mathcal{H}_{\mathrm{eff}}$ and
dissipator $\mathcal{D}$ have contributions from
all jump operators
$\mathcal{J}\in\{\hat{\sigma}_{j}^{-}\sqrt{\Gamma_{\mathrm{ud}}/2},\hat{\sigma}_{j}^{+}\sqrt{\Gamma_{\mathrm{du}}/2},\hat{\sigma}_{j}^{z}\sqrt{\Gamma_{\mathrm{el}}/8}:1\leq
j\leq\mathcal{N}\}$,
\begin{equation}
\label{meq2}
\mathcal{H}_{\mathrm{eff}}=\mathcal{H}-i\sum_{\mathrm{all}~\mathcal{J}}\mathcal{J}^{\dagger}\mathcal{J}^{},~~~~~\mathcal{D}(\rho)=2\sum_{\mathrm{all}~\mathcal{J}} \mathcal{J}^{}\rho\mathcal{J}^{\dagger},
\end{equation}
where $\hat{\sigma}_j^{\pm}=\frac{1}{2}(\hat{\sigma}_j^x\pm i\hat{\sigma}_j^y$).
The jump operators $\hat{\sigma}^-$, $\hat{\sigma}^+$, and
$\hat{\sigma}^z$ give rise to spontaneous de-excitation, spontaneous excitation, and elastic dephasing, respectively (see
Fig. \ref{Fig1}).  We refer to the spin-changing processes ($\hat{\sigma}^\pm$) as
Raman decoherence, and the spin-preserving processes
($\hat{\sigma}^z$) as Rayleigh decoherence.  In what follows we assume
an initially pure and uncorrelated density operator $\rho(0)=|\psi(0)\rangle\langle\psi(0)|$, with
$|\psi(0)\rangle=\bigotimes_{j}\sum_{\sigma_j}f_j(\sigma_j)|\sigma_j\rangle$.
Here $\sigma_j=\pm1$ are the eigenvalues of $\hat{\sigma}^z_j$,
$f_{j}(1)=\cos(\theta_j/2)e^{i\varphi_j/2}$, and
$f_{j}(-1)=\sin(\theta_j/2)e^{-i\varphi_j/2}$, for arbitrary
$\theta_j$ and $\varphi_j$.

Our approach to the solution of Eq. (\ref{meq}) for the chosen initial
conditions and arbitrary $\{\Gamma_{\mathrm{ud}}$,
$\Gamma_{\mathrm{du}},\Gamma_{\mathrm{el}}\}$ is based on the
quantum trajectories method \cite{QuantumTrajectories}, in which $|\psi(0)\rangle$ is
time-evolved with the effective Hamiltonian
\begin{equation}
\label{heff}
\mathcal{H}_{\mathrm{eff}}=\mathcal{H}-\frac{i}{2}\sum_{j}\left(\frac{\Gamma_{\mathrm{r}}}{2}+\frac{\Gamma_{\mathrm{el}}}{4}+2\gamma\hat{\sigma}^{z}_{j}\right),
\end{equation}
and the dynamics are interspersed with stochastic applications of the jump
operators.  In Eq. (\ref{heff}) we have defined
$\Gamma_{\mathrm{r}}=\Gamma_{\mathrm{ud}}+\Gamma_{\mathrm{du}}$ and
$\gamma=\frac{1}{4}(\Gamma_{\mathrm{ud}}-\Gamma_{\mathrm{du}})$.
According to the standard prescription \cite{QuantumTrajectories}, a particular
trajectory consists of a set of jump times $\{t_1,t_2,...\}$, which are
selected by equating the norm of the wave-function to a random number
uniformly distributed between 0 and 1.  Choosing which jump operator to
apply at each time requires calculating all expectation values
$\langle\mathcal{J}^{\dagger}\mathcal{J}\rangle$.  Because $\mathcal{H}$ is Hermitian and commutes with all products $\mathcal{J}^{\dagger}\mathcal{J}$, it has no
effect on the selection of the jumps, which can therefore be obtained for each spin independently (since the anti-Hermitian part
of $\mathcal{H}_{\mathrm{eff}}$ does not couple different spins).  With the
jump times and jump operators in hand, we define a string of $n_j$ (time-labeled)
jump operators on site $j$ as
$\hat{\mathcal{Q}}_j=\mathcal{J}_j^1(t_j^1)\times\dots\times\mathcal{J}_j^{n_j}(t_j^{n_j})$.
The time evolution of the wavefunction along a trajectory is then
\begin{equation}
\label{timeorder}
|\psi(t)\rangle=\mathcal{T}(e^{-i\mathcal{H}_{\mathrm{eff}}t}\prod_j\hat{\mathcal{Q}}_j)|\psi(0)\rangle,
\end{equation}
where the time-ordering operator $\mathcal{T}$ enforces that
the jump operators are interspersed in the time evolution according to
their time labels.

\begin{figure}[t!]
\includegraphics[width=8.6cm]{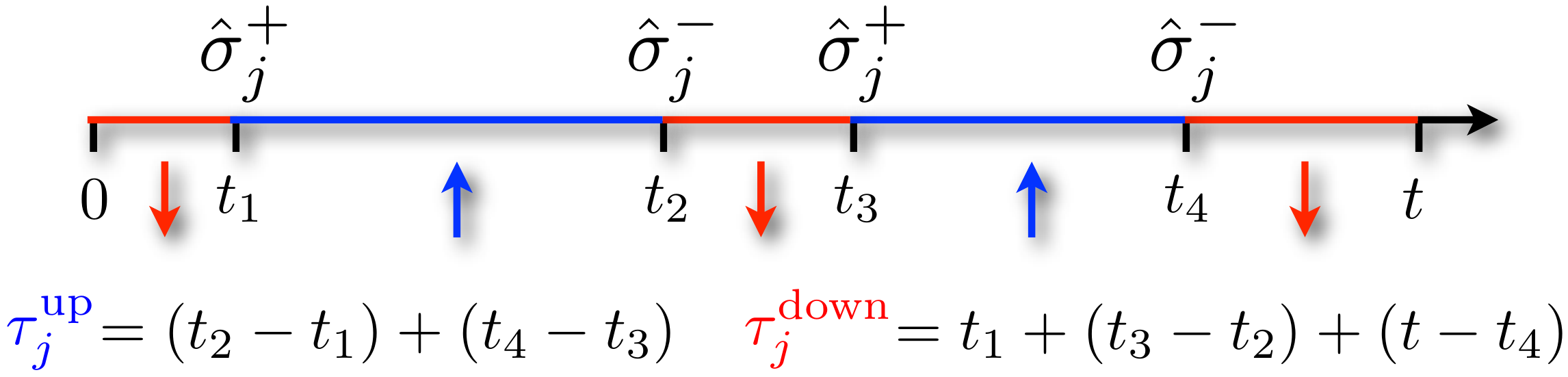}
\vspace{0.0cm}
\caption{ (Color online) A series of Raman flips of the spin on
  site $j$ can be formally accounted for as a magnetic field of
  strength $2J_{jk}/\mathcal{N}$ acting for a period of time
  $\tau_j^{\mathrm{up}}-\tau_j^{\mathrm{down}}$.  In the notation
  defined below, this series of jumps is represented by the operator $\hat{\mathcal{Q}}_j\propto\hat{\sigma}^{+}_j(t_1)\hat{\sigma}^{-}_j (t_2)\hat{\sigma}^{+}_j (t_3)\hat{\sigma}^{-}_j (t_4)$.}
 \label{Fig2}
\end{figure}

The time-ordering of the Rayleigh jumps can be ignored: Because $[\hat{\sigma}^z_j,\mathcal{H}_{\mathrm{eff}}]=0$ and
$[\hat{\sigma}_j^z,\hat{\sigma}_j^{\pm}]=\pm2\hat{\sigma}_{j}^{\pm}$, all
Rayleigh jumps can be evaluated at $t=0$ (their commutation with Raman
jumps affects only the overall sign of the wavefunction).  To the contrary, the Raman jumps do not commute with
$\mathcal{H}_{\mathrm{eff}}$, and their time ordering cannot be so
easily accounted for.  However, imagine that spin $j$ undergoes a
single Raman jump, created by applying $\hat{\sigma}_{j}^{+}$ at
time $t$.  This jump operator not only flips spin $j$ into the up
position, but also removes all parts of the wavefunction in which spin
$j$ pointed up immediately before time $t$.  Hence it is as this spin
pointed down before time $t$, and up after time $t$.  Since spin $j$ is always in an eigenstate of $\hat{\sigma}_j^z$, it is a spectator to
the Ising dynamics, but it does influence the other spins via the
Ising coupling $J_{jk}$; formally it acts on spin $k$ as an inhomogeneous magnetic field of
strength $2J_{jk}/\mathcal{N}$ that pointed down before $t$ and up after $t$.
For a spin on site $j$ that undergoes multiple Raman processes, the same reasoning allows us to treat it as a field of
strength $2J_{jk}/\mathcal{N}$ that acts for a period of time
$\tau_j=(\tau_j^{\mathrm{up}}-\tau_j^{\mathrm{down}})$, where
$\tau_j^{\mathrm{up(down)}}$ is the total amount of time that spin $j$ spends
pointing up(down) along $z$ (see Fig.
\ref{Fig2}).  Hence we are free to evaluate
all of the jump operators at $t=0$ to give
$|\tilde{\psi}\rangle=\prod_j\hat{\mathcal{Q}}_j|\psi\rangle$, thus ignoring the
time-ordering in Eq. (\ref{timeorder}), so long as we evolve
$|\tilde{\psi}\rangle$ with a modified time-evolution operator
\begin{equation}
\label{ueffective}
\mathcal{U}=\exp\left[-it\left(\mathcal{H}^{'}+\sum_{j}(\eta_j-i\gamma)\hat{\sigma}^z_j.\right)\right].
\end{equation}
Here $\eta_j=\frac{1}{\mathcal{N}t}\sum_k J_{jk}\tau_k$ accounts for the
magnetic field of all Raman-flipped ions,  $\mathcal{H}^{'}$ is obtained
from $\mathcal{H}$ by ignoring the spin-spin couplings to spins that
have undergone Raman jumps, and $\gamma$ accounts for the non-Hermitian part of $\mathcal{H}_{\mathrm{eff}}$ that is \emph{not}
proportional to the identity operator.

The expectation value of an arbitrary operator $\hat{\mathcal{O}}$
at the end of a particular quantum trajectory is therefore given by
$\langle{\hat{\mathcal{O}}\rangle}=\langle\tilde{\psi}|\mathcal{U}^{\dagger}\hat{\mathcal{O}}\mathcal{U}|\tilde{\psi}\rangle/\langle\tilde{\psi}|\mathcal{U}^{\dagger}\mathcal{U}|\tilde{\psi}\rangle$, and formally
taking the average over all trajectories (denoted by an overbar) we
have $\trace[\rho\hat{\mathcal{O}}]=\overline{\langle\hat{\mathcal{O}}\rangle}$.  Along with Eq. (\ref{ueffective}), these constitute a formal
solution for the dynamics of any observable.  We now proceed to derive closed-form expressions for the transverse spin-length and spin-spin
correlation functions---which have not been derived previously even in
the \textit{absence} of decoherence.  These are the central result of this paper.

\emph{Transverse-spin length and correlation functions.}---Relaxation of the transverse spin-length in an Ising-type spin model is a canonical example of equilibration in
a closed quantum system \cite{emch,radin,kastner}.  In the present model, such relaxation occurs due to a combination of the
proliferation of quantum fluctuations \emph{and} the equilibration
with the environment (decoherence).  Our theoretical treatment allows for both effects to be treated
simultaneously, and therefore, in principle, the disentangling of
these two physically different (but consequentially similar)
processes.  While such relaxation does not directly indicate quantum correlations or
entanglement, in the absence of decoherence it nevertheless is entirely due to the buildup of quantum
correlations---at the mean-field level coherent
relaxation is absent. The correlations that develop
during the dynamics can be understood in more detail by looking at
two-spin correlation functions---for instance, these furnish a complete
description of spin squeezing \cite{ueda}.  In the
absence of decoherence and for all-to-all coupling ($\zeta=0$), it is well known \cite{ueda} that the transverse spin component revives at a time
$\tau_{\mathrm{r}}=\mathcal{N}\pi\hbar/(2 J)$, and that
a highly-entangled macroscopic-superposition state (MSS) appears at
time $\tau_{\mathrm{r}}/2$.  This state is characterized by vanishing spin
length but maximum transverse spin fluctuations, and our solution
can be used to assess the robustness of such fluctuations against
decoherence.

To calculate the transverse spin-length along a particular trajectory, we assign the discrete-valued variables
$\mathcal{R}_j$ and $\mathcal{F}_j$ to each lattice
site, which count the number of Raman jumps and Rayleigh jumps,
respectively.  As we have discussed earlier, all jump operators can be applied
at $t=0$, and therefore specifying the
$\{\mathcal{R}_j,\mathcal{F}_j,\tau_j\}$ on each site fully determines
transverse spin-length along that trajectory.  The transverse spin component in
direction $\varphi$ and on site $j$ is given simply in terms of the
spin-raising operator on that site, $\langle
S_j^{\varphi}\rangle=\cos\varphi\langle S^x_j\rangle+\sin\varphi
\langle S^y_j\rangle=\mathrm{Re}[e^{-i\varphi}\langle\hat{\sigma}_j^+\rangle]$,
and generalizing solutions obtained in Ref. \cite{hazzard,emch} we find,
\begin{equation}
\label{sigmaplus}
\langle\hat{\sigma}_j^{+}\rangle=\frac{\alpha_j\beta_j\sin\theta_je^{i\varphi_j}}{2g_j(2\gamma
      t)}\prod_{k\neq j}e^{\frac{2iJ_{kj}
  \tau_{k}}{\mathcal{N}}}\frac{g_k[2\alpha_k
    t(\gamma-iJ_{jk}/\mathcal{N})]}{g_k(2\gamma t\alpha_k)}.
\end{equation}
Here $g_j(x)=\sum_{\sigma}|f_j(\sigma)|^2e^{-\sigma x}$,
$\alpha_j=\delta_{\mathcal{R}_j,0}$ ($\delta$ being the
Kronecker-delta symbol), $\beta_j=(-1)^{\mathcal{F}_j}$, and the
details of the calculation are given in the Supplementary material.
Defining a function $\mathcal{P}(\mathcal{R},\mathcal{F},\mathcal{\tau})$ that determines
the probability distribution of these variables on a given lattice
site, we have
\begin{equation}
\label{formalanswer}
\overline{\langle\hat{\sigma}_j^{+}\rangle}=\!\!\!\sum_{\mathrm{all}~\mathcal{R}}\sum_{\mathrm{all}~\mathcal{F}}\int\!\!\dots\int\!
\prod_{k}d\tau_{k}\mathcal{P}(\mathcal{R}_k,\mathcal{F}_k,\tau_k) \langle\hat{\sigma}_j^{+}\rangle.
\end{equation}

Equation (\ref{formalanswer}) constitutes a formal solution for
$\overline{\langle\hat{\sigma}_j^{+}\rangle}$, and it can always be evaluated efficiently by
averaging $\langle\hat{\sigma}_j^{+}\rangle$ over stochastically generated trajectories.  However, because the noise is uncorrelated from site to site, and
accordingly the expression inside the product of Eq. (\ref{sigmaplus})
depends only on local stochastic variables, these sums and integrals
factor into $\mathcal{N}$ independent sets (each over the
three stochastic variables), admitting closed form expressions.  At
this point, to avoid unnecessary complications in the ensuing
expressions, we will take our initial state to point along the $x$
axis ($\theta=\pi/2$, $\varphi=0$), but our results easily generalize \cite{inprep}.  In the Supplement we explain
how to evaluate these sums and integrals, and here we simply quote the
result.  Defining
\begin{equation}
\Phi(J,t)=e^{-\lambda t}\left[\cos\left(t\sqrt{s^2-r}\right)+\lambda t\sinc\left(t\sqrt{s^2-r}\right)\right], \nonumber
\end{equation}
with $\lambda=\Gamma_{\mathrm{r}}/2$, $s=2i\gamma+2J/\mathcal{N}$ and
$r=\Gamma_{\mathrm{ud}}\Gamma_{\mathrm{du}}$, we find
\begin{equation}
\label{tsl}
\overline{\langle\hat{\sigma}_j^{+}\rangle}=\frac{1}{2}e^{-\Gamma t}\prod_{k\neq
  j}\Phi(J_{jk},t),
\end{equation}
where the total decoherence rate is defined
$\Gamma=\frac{1}{2}(\Gamma_{\mathrm{r}}+\Gamma_{\mathrm{el}})$ \footnote{Despite
  superficial appearances, notice that $\Phi(0,t)=1$ (and not
  $e^{-\Gamma_{\mathrm{r}}t/2}$), justifying the designation of
  $\Gamma$ as the total decay rate, as in \cite{uys}.}.  Similar
calculations to those described above yield spin-spin correlation functions
\begin{eqnarray}
\label{correlations1}
\overline{\langle
\hat{\sigma}^{\mu}_j\hat{\sigma}^{\nu}_k\rangle}&=&\frac{1}{4}e^{-2\Gamma
  t}\prod_{l\notin\{j,k\}}\Phi(\mu J_{jl}+\nu J_{kl},t) \\
\label{correlations2}\overline{\langle
\hat{\sigma}^{\mu}_j\hat{\sigma}^z_k\rangle}&=&\frac{1}{2}e^{-\Gamma
  t}\Psi(\mu J_{jk},t)\prod_{l\notin\{j,k\}}\Phi(\mu J_{jl},t),
\end{eqnarray}
with $\Psi(J,t)=e^{-\lambda t}(is-2\gamma)t\sinc(t\sqrt{s^2-r})$ and $\mu,\nu=\pm$.
\begin{figure*}[t!!]
\centering
\subfiguretopcaptrue
\subfigure[][]{
\label{Fig3a}
\includegraphics[height=3.78 cm]{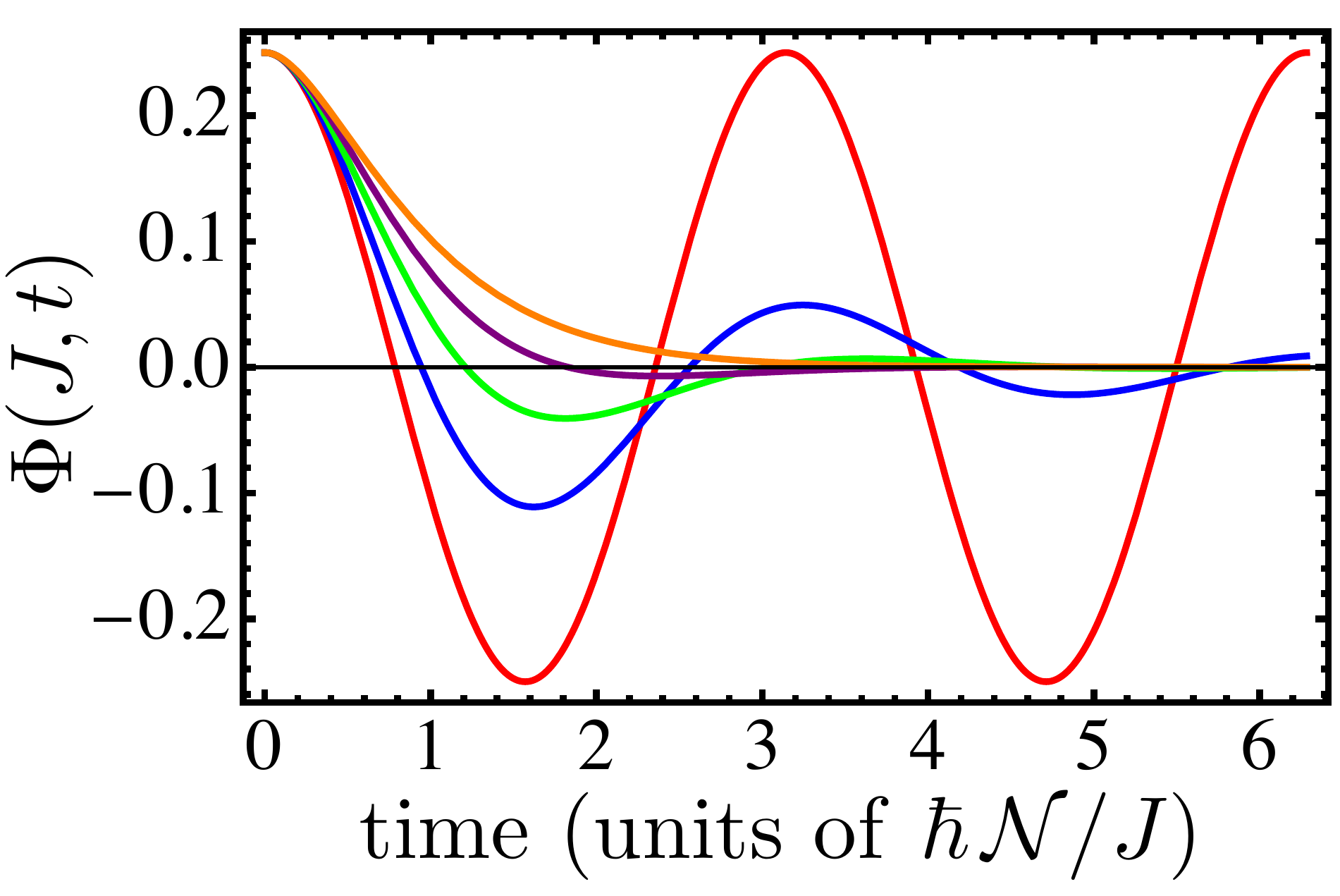}
}
\subfigure[][]{
\label{Fig3b}
\includegraphics[height=3.78 cm]{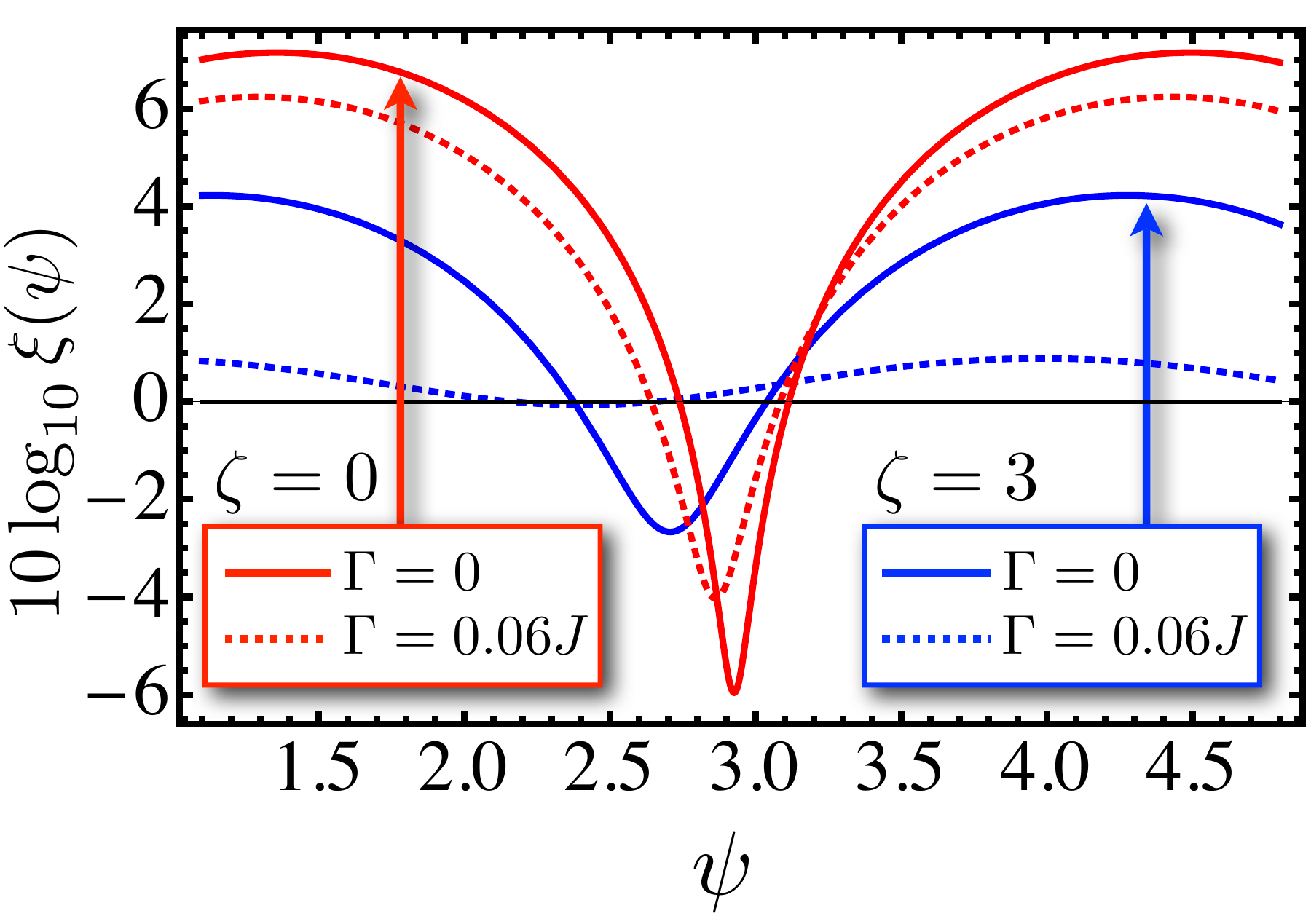}
}
\subfigure[][]{
\label{Fig3c}
\includegraphics[height=3.78 cm]{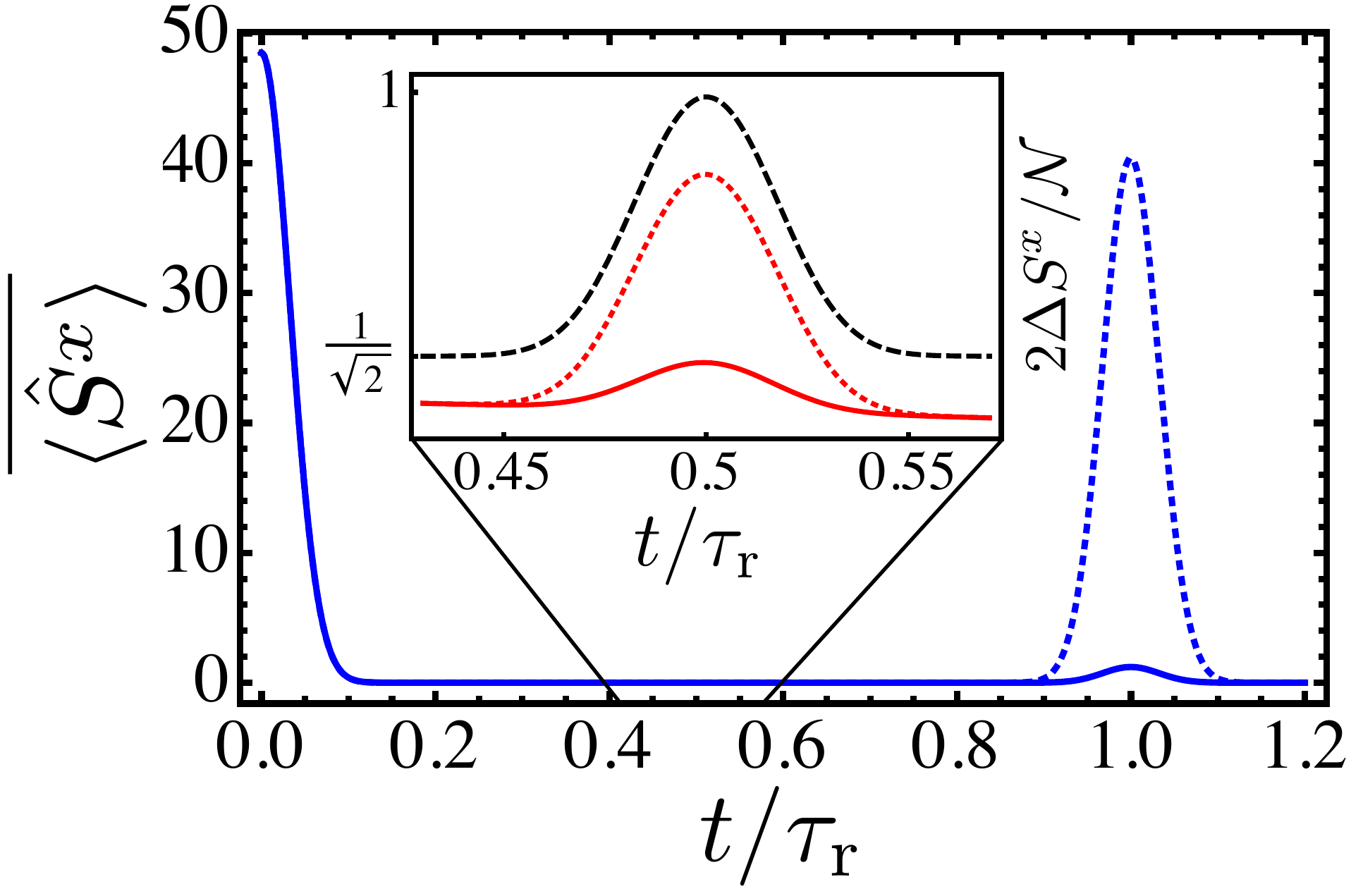}
}
\caption{ (Color online) (a) Plots of $\Phi(J,t)$ for $\gamma=0$ and $\Gamma_{\mathrm{r}}/\Gamma_{\mathrm{r}}^{\mathrm{c}}\in\{0,1/4,1/2,3/4,1\}$, showing a
  transition from oscillatory to damped behavior. (b) Example of how spin squeezing is affected by decoherence (dashed
  lines are $\Gamma=0.06 J$, solid lines are for $\Gamma=0$) for
  long-ranged (red, $\zeta=0$) and short-ranged (blue, $\zeta=3$)
  interactions. (c) Transverse spin relaxation and revivals for $\zeta=0$, with parameters corresponding to expected experimental
  capabilities in \cite{britton} (blue
  solid line).  The dotted blue line
  is obtained by treating decoherence at the single-particle level,
  and underestimates the detrimental effect of Raman
  decoherence by about a factor of 35.  Inset: transverse-spin fluctuations peaking
  at time $\tau_\mathrm{r}/2$.  Experimental parameters, exact
  treatment (red solid line); experimental parameters,
  single-particle treatment (red-dotted line); no decoherence (black
  dashed line).}
 \label{Fig3}
\end{figure*}
These correlation functions, along with similar ones obtained by interchange of the site indices, completely
determine the spin-spin correlations.  Each instance of
$\hat{\sigma}^x$ or $\hat{\sigma}^y$ in an observable generates an
overall multiplicative factor of $e^{-\Gamma t}$; this is the effect
of decoherence at the single-particle level.  The structure of
$\Phi(J,t)$ captures the effect of decoherence on the many-body
physics, and could not have been deduced without our exact treatment.

We note that for $\gamma=0$ the function $\Phi(J,t)$ undergoes a
qualitative transition from oscillatory ($s^2>r$) to damped ($s^2<r$) behavior when
$\Gamma_r=\Gamma_{\mathrm{r}}^{\mathrm{c}}\equiv4J/\mathcal{N}$ [Fig. \ref{Fig3a}].  Therefore, when there is only one
coupling strength $J$, for instance in the case of all-to-all or nearest-neighbor
couplings, dynamics of the transverse spin-length and
correlation functions undergoes the same transition.  More generally,
for nonzero $\gamma$ and any $\zeta$, correlation functions exhibit a
series of Hopf bifurcations as $\Gamma_{\mathrm{r}}$ is tuned.  We note that the factors of $e^{-\Gamma t}$
in Eqs. (\ref{tsl}-\ref{correlations2}) may make this transition difficult to
observe experimentally, since the correlations are rapidly
suppressed at the critical $\Gamma_{\mathrm{r}}^{\mathrm{c}}$.

\emph{Application to a trapped ion quantum simulator}.---Trapped
ion systems can simulate the Hamiltonian in Eq. (\ref{hamiltonian}), and can accurately measure the decoherence rates
$\Gamma_{\mathrm{el}}$, $\Gamma_{\mathrm{ud}}$, and
$\Gamma_{\mathrm{du}}$.  We note that both the Born-Markov
approximation and the assumption of uncorrelated decoherence processes
are extremely well justified for trapped ion systems \cite{uys}.  Sample averaged spin-length and spin-spin correlation
functions are easily measured in these experiments by looking at the length (and its shot-to-shot fluctuations) of various
projections of the Bloch vector.  In the trapped ion experiments
discussed in Ref. \cite{britton}, when $\zeta=0$ the time scale at which
quantum correlations become important for these observables, $\tau_{c}$, scales with some power of the
ion number: $\tau_c\sim\mathcal{N}^{1/3}$ for spin-squeezing, $\sim\mathcal{N}^{1/2}$ for transverse-spin
relaxation, and $\sim\mathcal{N}$ for the creation of MSS's.  Taking $\mathcal{N}=100$ and $\Gamma=0.06J$, as
is typical in that experiment, we expect the proper incorporation of
decoherence to be quantitatively important even for spin-squeezing,
despite it being a relatively short-time indication of entanglement.

Equations (\ref{correlations1},\ref{correlations2}) allow us to exactly calculate the effect of
decoherence and the finite range of interactions on the maximum spin
squeezing achievable in experiment.  Figure \ref{Fig3b} shows the
expected squeezing and anti-squeezing as a function of angle
$\psi$ in the $yz$ plane for $\zeta=0,3$, with $\Gamma=0.06J$ and
$\Gamma_{\mathrm{el}}=8\Gamma_{\mathrm{ud}}=8\Gamma_{\mathrm{du}}$ (typical experimental
numbers in \cite{britton}).  The effects of decoherence are more pronounced for shorter-range
interactions due to the longer time scales for maximal
squeezing.  For this calculation the spins are assumed to be initialized (prior to the Ising dynamics) in a coherent state pointing
along the $x$-axis, and we define the spin squeezing parameter
\begin{equation}
\xi(\psi)=\frac{\sqrt{\mathcal{N}}\Delta S^{\psi}}{\overline{\langle \hat{S}^x\rangle}},
\end{equation}
with $\Delta A=\sqrt{\overline{\langle
    \hat{A}^2\rangle}-\overline{\langle\hat{A}\rangle}^2}$,
$\hat{S}^{\psi}=\frac{1}{2}\sum_{j}\hat{\sigma}_j^{\psi}$, and
$\hat{\sigma}_j^{\psi}=\sin(\psi)\hat{\sigma}_j^y+\cos(\psi)\hat{\sigma}_j^z$.
In Fig. \ref{Fig3b}, $\xi$ has been
calculated at the time when maximal squeezing occurs.

For the current experimental parameters there is essentially no spin
revival, and no indication of a MSS at $\tau_{\mathrm{r}}/2$.  However, assuming $97$ ions
and expected improvements in the experiment \cite{britton} (a roughly
$50$-fold increase in the ratio $J/\Gamma$), in Fig. \ref{Fig3c} we
show that transverse spin revivals begin to appear.  In that figure, the dotted line is obtained by treating the
decoherence at the single-particle level, which amounts to attaching a decaying
exponential $e^{-\Gamma t}$ to the operators $\hat{\sigma}^x$ and
$\hat{\sigma}^y$.  We note that this treatment would be exact if the
decoherence were only of the Rayleigh type ($\Gamma_{\mathrm{r}}=0$).  The solid line is the full solution from
Eq. (\ref{tsl}); the large ($\sim35$-fold) discrepancy between the
single-particle and exact results indicates that our theory will be
essential for understanding the behavior of this experiment.  In the inset of Fig. \ref{Fig3c} we plot the transverse
spin fluctuations $\Delta S^x$ at times near $\tau_{\mathrm{r}}/2$.  The
peak in these fluctuations (which would achieve unity in the
absence of decoherence) is a result of strong transverse-spin
correlations in an emerging MSS, indicating that the expected
improvements to the experiments will bring within reach the production
of MSS's of $\sim100$ ions in the near future.

\emph{Conclusions.}---We have exactly solved the non-equilibrium
dynamics of arbitrary Ising models in the presence of single-particle decoherence.  These calculations provide a rare glimpse into
the exact structure of relaxation dynamics in open and
strongly interacting quantum systems.  They will allow for
quantitative insights into experiments studying interacting
spin-systems that are not perfectly isolated from the environment.  In
particular, they provide a rigorous foundation for
benchmarking trapped-ion quantum simulators when interactions
are strong and decoherence is non-negligible on experimental timescales.

We gratefully acknowledge Joe Britton, Brian Sawyer, Salvatore
Manmana, Michael Kastner, Dominic Meiser, and Murray Holland for helpful discussions.  This work was supported by NIST, the NSF (PIF and
PFC grants), AFOSR and ARO individual investigator awards, and the ARO
with funding from the DARPA-OLE program.  K.H. thanks the NRC for
support.  During the preparation of
this manuscript, we were informed of simultaneous calculation of
spin-spin correlation functions for quantum Ising models in the
absence of decoherence \cite{kastner2}. This manuscript is the contribution
of NIST and is not subject to U.S. copyright.

\setcounter{equation}{0}
\renewcommand{\theequation}{S\arabic{equation}}
\appendix
\begin{widetext}
\section{Expressions for spin length and correlation functions along a
  single trajectory}

As in the main text, we write the state of a single spin
as $\sum_{\sigma_j^z}f_j(\sigma_j^z)|\sigma_j^z\rangle$, where $\sigma_j^z$ is an index that takes on the eigenvalues of the
operator $\hat{\sigma}_j^z$, $f_j(1)=\cos\left(\theta_j/2\right)e^{i\varphi_j/2}$, and
$f_j(-1)=\sin\left(\theta_j/2\right)e^{-i\varphi_j/2}$.  The initial state of
the entire system is taken to be a direct product of states for each
individual spin
\begin{eqnarray}
|\psi(0)\rangle&\equiv&\bigotimes_{j}\sum_{\sigma_j}f_j(\sigma_j)|\sigma_j\rangle
\\
&=&\sum_{\sigma_1^z,\dots,\sigma^z_{\mathcal{N}}}f_1(\sigma_1^z)\times\dots\times f_{\mathcal{N}}(\sigma_{\mathcal{N}}^z)|\sigma_1^z,\dots,\sigma_{\mathcal{N}}^z\rangle.\nonumber
\end{eqnarray}
As discussed in the main text, evaluating all of the Rayleigh jumps at $t=0$ can be accomplished by
changing $\varphi_j\rightarrow\varphi_j+\pi\mathcal{F}_j$, which rotates the
spin on site $j$ by an angle $\pi$ if it has undergone an odd number
of Rayleigh jumps.  Raman jumps can be incorporated at $t=0$ by
setting $\theta_j=0(\pi)$ if the final Raman jump on site $j$ was
$\hat{\sigma}^{+}_j(\hat{\sigma}_j^{-})$.  We therefore define
$\tilde{f}_j(\sigma_j^z)$ to be the modification of $f_j(\sigma_j^z)$
under those transformations.  We also note that spins having undergone one or more Raman
jumps are treated as an effective magnetic field, and not included in
the spin-spin coupling term of the Hamiltonian.  This is accomplished
by changing $J_{ij}\rightarrow\alpha_i\alpha_j J_{ij}$ (with $\alpha_j=\delta_{\mathcal{R}_j,0}$) in the
Hamiltonian and including an effective magnetic field.  Therefore, we can
write the (time-dependent) wavefunction $|\tilde{\psi}(t)\rangle$ evolving under
$\mathcal{U}$ [as defined in Eq. (\ref{ueffective}) of the manuscript] as
\begin{equation}
|\tilde{\psi}(t)\rangle=\sum_{\sigma_1^z,\dots,\sigma_{\mathcal{N}}^z}\exp\left[-it\left(\frac{1}{\mathcal{N}}\sum_{i>j}\alpha_i\alpha_jJ_{ij}\sigma_i^z\sigma_j^z+\sum_{j}(\eta_j-i\gamma)\sigma_j^z\right)\right]\tilde{f}_1(\sigma_1^z)\times\dots\times \tilde{f}_{\mathcal{N}}(\sigma_{\mathcal{N}}^z)|\sigma_1^z,\dots,\sigma_{\mathcal{N}}^z\rangle.
\label{wf}
\end{equation}
In order to calculate the transverse spin-length we will first
calculate
$\langle\hat{\sigma}_j^{+}\rangle=\langle\tilde{\psi}(t)|\hat{\sigma}^{+}_j|\tilde{\psi}(t)\rangle/\langle\tilde{\psi}(t)|\tilde{\psi}(t)\rangle$,
and at the end obtain $\langle\hat{S}^x\rangle=\mathrm{Re}\left(\sum_{j}\langle\hat{\sigma}_j^{+}\rangle\right)$.

Let's imagine, in particular, calculating $\langle\tilde{\psi}(t)|\hat{\sigma}_1^+|\tilde{\psi}(t)\rangle$
(there is nothing special about the first spin, this just makes the
notation in what follows less confusing).  Because the wavefunction
enters twice, this would involve two sums like the one in
Eq. (\ref{wf}), over $\sigma_j^z$ and $\sigma_j^{'z}$, but very few
terms survive: We need $\sigma_1^z=-1$, $\sigma_1^{'z}=1$, and for all
$j\neq1$ we must have $\sigma_{j}^z=\sigma_j^{'z}$, so the matrix
element is given by
\begin{equation}
\langle\tilde{\psi}(t)|\hat{\sigma}_1^+|\tilde{\psi}(t)\rangle=\tilde{f}_1 ^{*} (1)\tilde{f}^{}_1(-1)\sum_{\sigma_2^z,\dots,\sigma_{\mathcal{N}}^z}|\tilde{f}_2(\sigma_2^z)|^2\times\dots\times |\tilde{f}_{\mathcal{N}}(\sigma_{\mathcal{N}}^z)|^2\exp\left[2it\left(\eta_1+\sum_{j=2}^{\mathcal{N}}\frac{1}{\mathcal{N}}\alpha_1\alpha_jJ_{1j}\sigma_j^z+i\alpha_j\gamma\sigma_j^z\right)\right].
\end{equation}
If $\alpha_{j}=0$, the $j^{\mathrm{th}}$ spin always has a well defined value of $\sigma_j^z$ and the
choice to include the term $\gamma\sigma_j^z$ in the exponentiated sum
(or not) affects only the overall normalization of the wavefunction.
By multiplying $\gamma$ by $\alpha_j$ (in the exponentiated sum), we
have chosen to not include the term $\gamma\sigma_j^z$, and this is properly accounted for when
normalizing this expectation value below.  In order to obtain
$\langle\hat{\sigma}_1^{+}\rangle$ we must divide by the (non-conserved)
normalization of the wavefunction
$\langle\tilde{\psi}(t)|\tilde{\psi}(t)\rangle$.  Defining
$g_{j}(x)=\sum_{\sigma}|f_{j}(\sigma)|^2e^{-\sigma x}$ (as in
the manuscript), we obtain
\begin{equation}
\langle\hat{\sigma}_1^{+}\rangle=\frac{\tilde{f}_1 ^{*}
  (1)\tilde{f}^{}_1(-1)}{g_1(2\gamma
  t)}\sum_{\sigma_2^z,\dots,\sigma_{\mathcal{N}}^z}\frac{|\tilde{f}_2(\sigma_2^z)|^2}{g_2(2\alpha_2\gamma
  t)}\times\dots\times
\frac{|\tilde{f}_{\mathcal{N}}(\sigma_{\mathcal{N}}^z)|^2}{g_{\mathcal{N}}(2\alpha_\mathcal{N}\gamma
  t)}\exp\left[2it\left(\eta_1+\sum_{j=2}^{\mathcal{N}}\frac{1}{\mathcal{N}}\alpha_1\alpha_jJ_{1j}\sigma_j^z+i\alpha_j\gamma\sigma_j^z\right)\right].
\label{wheretoinsertz}
\end{equation}
This expression can be simplified by making the following set of observations: (1) $\tilde{f}_1
^{*}
(1)\tilde{f}^{}_1(-1)=\alpha_1\beta_1e^{i\varphi_1}\sin(\theta_1)/2$
[as in the text, $\beta_j=(-1)^{\mathcal{F}_j}$], (2) the
$\alpha_1$ in the exponent is irrelevant, because if it takes the
value $0$ the entire expression
for$\langle\tilde{\psi}(t)|\hat{\sigma}_1^+|\tilde{\psi}(t)\rangle$
vanishes, and (3) the summand factorizes into a product where each term contains only local (i.e. on a single site) variables, and hence the sum of products can be exchanged for a product of sums.  Taking (1-3) into account we obtain
\begin{eqnarray}
\langle\hat{\sigma}_1^{+}\rangle&=&\frac{\sin(\theta_1)e^{i\varphi_1}}{2g_1(2\gamma
  t)}\alpha_1\beta_1\prod_{j=2}^{\mathcal{N}}\left(\sum_{\sigma^z_j}\frac{|\tilde{f}_j(\sigma_j^z)|^2}{g_j(2\alpha_j\gamma
    t)}\exp\left[2it\alpha_j
    \sigma_j^z\left(J_{1j}/\mathcal{N}+i\gamma\right)\right]e^{2iJ_{1j}\tau_j/\mathcal{N}}\right). \\
&=&\frac{\sin(\theta_1)e^{i\varphi_1}}{2g_1(2\gamma
  t)}\alpha_1\beta_1\prod_{j=2}^{\mathcal{N}}\left(e^{2iJ_{1j}\tau_j/\mathcal{N}}\frac{g_j[2\alpha_j t(\gamma-iJ_{1j}/\mathcal{N})]}{g_j(2\gamma
    t\alpha_j)}\right).
\end{eqnarray}
Therefore, we can write
\begin{equation}
\label{SuppSx}
\langle\hat{S}^x\rangle=\Re\sum_{j=1}^{\mathcal{N}}\left[\frac{\sin(\theta_j)e^{i\varphi_1}}{2g_j(2\gamma
  t)}\alpha_j\beta_j\prod_{k\neq j}\left(e^{2iJ_{jk}\tau_k/\mathcal{N}}\frac{g_k[2\alpha_k t(\gamma-iJ_{jk}/\mathcal{N})]}{g_k(2\gamma
    t\alpha_k)}\right)\right].
\end{equation}

The calculation of correlation functions follows from extremely
similar considerations.  For instance, let's consider $\mathcal{G}_{jk}^{-+}\equiv\langle\hat{\sigma}_j^{-}\hat{\sigma}_k^{+}\rangle$.  In this case, the operators in the expectation value only connect two states if $-\sigma^{'z}_j=\sigma^z_j=1$, $\sigma^{'z}_{k}=-\sigma^z_k=1$, and $\sigma^{'z}_l=\sigma^z_l$ whenever $l\notin\{j,k\}$.  Therefore, much as before we have
\begin{equation}
\label{SuppCorrelation}
\mathcal{G}_{jk}^{-+}=\frac{\sin(\theta_j)\sin(\theta_k)e^{i(\varphi_k-\varphi_j)}}{4g_j(2\gamma
  t)g_k(2\gamma
  t)}\alpha_j\alpha_k\beta_j\beta_k\prod_{l\notin\{j,k\}}\left(e^{2i(J_{kl}-J_{jl})\tau_l/\mathcal{N}}\frac{g_l[2\alpha_l t(\gamma-i[J_{kl}-J_{jl}]/\mathcal{N})]}{g_l(2\gamma
    t\alpha_l)}\right).
\end{equation}
Computing correlation functions involving a single
$\hat{\sigma}^z$, such as $\mathcal{G}_{jk}^{z+}\equiv\langle\hat{\sigma}_j^{z}\hat{\sigma}_k^{+}\rangle$
can be achieved by inserting $\sigma_{j}^z$ into the sum in
Eq. (\ref{wheretoinsertz}), yielding
\begin{equation}
\mathcal{G}_{jk}^{z+}=\frac{\sin(\theta_k)e^{i\varphi_k}\alpha_k\beta_k}{2g_k(2\gamma
  t)}\left[\alpha_j\frac{\cos^2(\theta/2)e^{-2\gamma
      t}-\sin^2(\theta/2)e^{2\gamma t}}{g_j(2\gamma t)}+(1-\alpha_j)\kappa_j\right]\prod_{l\notin\{j,k\}}\left(e^{2iJ_{kl}\tau_l/\mathcal{N}}\frac{g_l[2\alpha_l t(\gamma-iJ_{kl}/\mathcal{N})]}{g_l(2\gamma
    t\alpha_l)}\right).
\end{equation}
Assuming one or more Raman flip occurred, the variable $\kappa_j$ takes on
the values $\pm1$ if the final Raman jump is $\hat{\sigma}_j^{\pm}$.

\section{Analytic evaluation of stochastic averaging of trajectories}

\noindent At this point in the calculation, for clarity of presentation, we set $\varphi_j=0$ and
$\theta_j=\pi/2$ (for all $j$), so all spins point along the $x$ axis at $t=0$.  Defining $\mathcal{P}(\mathcal{R},\mathcal{F},\tau)$ to be the probability
distribution of the variables $\mathcal{R}$, $\mathcal{F}$, and $\tau$
on a single site (it is the same on every site), the trajectory averaged expectation value is given by
\begin{equation}
\overline{\langle\hat{\sigma}_j^{+}\rangle}=\sum_{\mathrm{all}~\mathcal{R}}\sum_{\mathrm{all}~\mathcal{F}}\int
d\tau_1\dots\int d\tau_{\mathcal{N}} \langle\hat{\sigma}_j^+\rangle\prod_{k}\mathcal{P}(\mathcal{R}_k,\mathcal{F}_k,\tau_k).
\end{equation}
To begin, we note that the probability distribution can be
decomposed as $\mathcal{P}(\mathcal{R},\mathcal{F},\tau)=\mathcal{P}_{\mathrm{el}}(\mathcal{F})\mathcal{P}_{\mathrm{r}}(\mathcal{R},\tau)$,
which is valid because the probability of Rayleigh jump is independent
of whether a Raman jump has occurred (and vice versa).  The occurrence
of random processes follows a Poissonian distribution, so
$\mathcal{P}_{\mathrm{el}}(\mathcal{F})=e^{-\Gamma_{\mathrm{el}}t/4}(\Gamma_{\mathrm{el}}t/4)^{\mathcal{F}}/\mathcal{F}!$,
and we have also calculated $\mathcal{P}_{\mathrm{r}}$.  The result depends
on whether $\mathcal{R}$ is even or odd (the proof is simple but requires
some careful reasoning, and we do not give it here).  We parameterize
the $\mathcal{R}$-odd solution by $\mu=(\mathcal{R}-1)/2$ (which will run over
all non-negative integers), and we parameterize the $\mathcal{R}$-even
solutions by $\mu=(\mathcal{R}-2)/2$ (once again running $\mu$ over all
nonnegative integers), and obtain
\begin{eqnarray}
\mathcal{P}_{\mathrm{r}}^{\mathrm{odd}}(\mu,\tau)&=&\frac{\Gamma_{\mathrm{r}}}{4}e^{-\Gamma_r t/2}\frac{(\Gamma_{\mathrm{ud}}\Gamma_{\mathrm{du}}/4)^{\mu}}{(\mu!)^2}e^{-2\tau\gamma}(t^2-\tau^2)^{\mu}\\
\mathcal{P}_{\mathrm{r}}^{\mathrm{even}}(\mu,\tau)&=&\frac{\Gamma_{\mathrm{ud}}\Gamma_{\mathrm{du}}t}{4}e^{-\Gamma_r t/2}\frac{(\Gamma_{\mathrm{ud}}\Gamma_{\mathrm{du}}/4)^{\mu}}{\mu!(\mu+1)!}e^{-2\tau\gamma}(t^2-\tau^2)^{\mu}.
\end{eqnarray}
By evaluating the sum
\begin{equation}
\sum_{\mathcal{F}=0}^{\infty}\mathcal{P}_{\mathrm{el}}(\mathcal{F})e^{i\pi\mathcal{F}}=e^{-\Gamma_{\mathrm{el}}t/2},
\end{equation}
we obtain
\begin{eqnarray}
\overline{\langle\hat{\sigma}_j^{+}\rangle}&=&\frac{e^{-\Gamma_{\mathrm{el}}t/2}}{2\cosh(2\gamma
  t)}\sum_{\mathcal{R}_1,\dots\mathcal{R}_{\mathcal{N}}}\int
d\tau_1\dots\int d\tau_{\mathcal{N}}\left[
\mathcal{P}(\mathcal{R}_j,\tau_j)\alpha_j\prod_{k\neq j}e^{2iJ_{jk}\tau_k/\mathcal{N}}\frac{\cosh[2\alpha_k t(\gamma-iJ_{jk}/\mathcal{N})]}{\cosh(2\gamma
    t\alpha_k)}\mathcal{P}_{\mathrm{r}}(\mathcal{R}_k,\tau_k) \right]\nonumber
\\
&=&\frac{e^{-\Gamma_{\mathrm{el}}t/2}}{2\cosh(2\gamma
  t)}\left[\sum_{\mathcal{R}_j}\int d\tau_j
  \alpha_j\mathcal{P}(\mathcal{R}_j,\tau_j)\right]\times\left[\prod_{j\neq
  k}\sum_{\mathcal{R}_k}\int d\tau_ke^{2iJ_{jk}\tau_k/\mathcal{N}}\frac{\cosh[2\alpha_k t(\gamma-iJ_{jk}/\mathcal{N})]}{\cosh(2\gamma
    t\alpha_k)}\mathcal{P}_{\mathrm{r}}(\mathcal{R}_k,\tau_k)\right].\nonumber
\end{eqnarray}
Because $\alpha_j$ gives 1 if there have not been any Raman flips at site
$j$ and 0 otherwise, the term in the first square bracket is just the probability that
there has been no Raman flip on site $j$, which is given by $e^{-\Gamma_{\mathrm{r}}t/2}\cosh(2\gamma t)$ [this comes from evolving the wavefunction of a single spin pointing along $x$ with the effective Hamiltonian Eq. (\ref{heff})].  Therefore we have
\begin{equation}
\overline{\langle\hat{\sigma}_j^{+}\rangle}=\frac{1}{2}e^{-\Gamma t}\times\left[\prod_{j\neq
  k}\sum_{\mathcal{R}_k}\int d\tau_ke^{2iJ_{jk}\tau_k/\mathcal{N}}\frac{\cosh[2\alpha_k t(\gamma-iJ_{jk}/\mathcal{N})]}{\cosh(2\gamma
    t\alpha_k)}\mathcal{P}_{\mathrm{r}}(\mathcal{R}_k,\tau_k)\right]
\end{equation}

Defining $s=2i\gamma+2J/\mathcal{N}$, the trick is now to evaluate the quantity
\begin{eqnarray}
\label{begin}
\Phi(J,t)&=&\sum_{\mathcal{R}}\int d\tau\mathcal{P}(\mathcal{R},\tau)
e^{2iJ\tau/\mathcal{N}}\frac{\cosh(is t\alpha)}{\cosh(2\gamma t \alpha)} \\
&=&e^{-\Gamma_{\mathrm{r}} t/2}\cosh(is t)+\sum_{\mathcal{R}=1}^{\infty}\int d\tau
\mathcal{P}(\mathcal{R},\tau)e^{2iJ\tau/\mathcal{N}.}\nonumber
\end{eqnarray}
The second equality follows from pulling off the $\mathcal{R}=0$ term in the sum,
which represents the probability of having no Raman flip (and so we
can set $\tau=0$ and $\alpha=1$ in this term).  The second term represents
the probability for any finite number of Raman flips, and hence we
must keep $\tau$ arbitrary but can set $\alpha=0$.  The integral over $\tau$ can be evaluated by using the identity
\begin{equation}
\int_{-t}^{t} d\tau(t^2-\tau^2)^{\mu}e^{-ix\tau}=(2t)^{\mu+1}\frac{j_{\mu}(x t) \mu!}{(x)^{\mu}},
\end{equation}
where $j$ is a spherical Bessel function.  The remaining sum over $\mathcal{R}$ can be recognized as a generating
function for the spherical Bessel functions (or a derivative thereof).  Defining parameters $\lambda=\Gamma_r/2$ and $r=\Gamma_{ud}\Gamma_{du}$, and functions
\begin{eqnarray}
F(x,y)&=&\mathrm{sinc}(\sqrt{x^2-y}) \\
G(x,y)&=&\frac{\cos(\sqrt{ x^2-y})-\cos(x)}{x}, 
\end{eqnarray}
we obtain
\begin{eqnarray}
\label{end}
\Phi(J,t)&=&e^{-\lambda t}\cos(s t)+\lambda t
e^{-\lambda t}F(s t,r t)+s t e^{-\lambda t}G(s t,r t)\\
&=&e^{-\lambda t}\left[\cos\left(t\sqrt{s^2-r}\right)+\lambda t\sinc\left(t\sqrt{s^2-r}\right)\right]
\end{eqnarray}
We can now write out the exact solution
\begin{equation}
\overline{\langle \hat{S}^x\rangle}=\frac{e^{-\Gamma t}}{2}\mathrm{Re}\sum_{j}\prod_{k\neq j}\Phi(J_{jk},t).
\end{equation}
Because Eqs. (\ref{SuppSx},\ref{SuppCorrelation}) have such a similar
structure, the stochastic averaging of correlation functions is almost
identical, and leads to the similar expressions given in the main text
[Eqs. (\ref{correlations1},\ref{correlations2})].

\end{widetext}

\end{document}